# Results from the Alghero Workshop on e$^+$e$^-$ in the 1-2 GeV range

C. Biscari

*LNF-INFN, Frascati, Italy*

In the post-LEP era the high energy frontier is reachable only by large scale collaborations and laboratories, while the present lepton colliders operating at low and intermediate energies are planning major upgrades aimed at increasing substantially their luminosity, in order to meet the continuous interest in precision physics. The lowest energies are now represented by DAΦNE (1.02 GeV cm, operating at Frascati), VEPP-2000 (from 1 to 2 GeV, in construction at Novosibirsk), CESR (presently operating at Cornell in the range between 3 to 12 GeV), BEPC (presently operating in Beijing in the range 2 to 5.6 GeV, foreseen luminosity upgrade). A workshop has been held to discuss both the physics issues and the strategies and problems towards higher luminosities in the energy range mainly but not only between 1 and 2 GeV. The workshop was intended to clarify which experiments are more appealing to the physics community, and which is the way to proceed for obtaining the necessary luminosity.

## 1. INTRODUCTION

The Workshop on "e$^+$e$^-$ in the 1-2 GeV range: Physics and Accelerator Prospects" was held from 10$^{th}$ to 13$^{th}$ September 2003 at Alghero, Sardinia [1]. It was organized by INFN and also sponsored as an ICFA Beam Dynamics Panel Miniworkshop, by the working group on High Luminosity e$^+$e$^-$ colliders.

Participation was shared between Experimental and Theoretical Physics and Accelerator Physics, attending the two different working groups on the Physics Case for the c.m. energy between 1 and 2 GeV and the luminosity issues at these energies.

A brief summary of the main arguments discussed in the workshop is here presented, with special emphasis on the accelerator group contributions. All the corresponding references can be found in the workshop proceedings [1].

## 2. PHYSICS CASE

The most appealing experiments among the discussed physics topics are the prospects for kaon physics and tests of fundamental symmetries through rare kaon decays. Here the impact of a very high luminosity future Φ-factory was extensively discussed also in relation with present and future fixed-target experiments. The requested luminosities are at least one order of magnitude higher than the present DAΦNE results.

Low energy spectroscopy and baryon form factors could be investigated in a new e$^+$e$^-$ collider at the c.m. energy around 2 GeV. Luminosity in the range of $10^{32}$cm$^{-2}$sec$^{-1}$ would greatly improve the present experimental accuracy.

In the field of hypernuclear spectroscopy, the recent start of data taking of the FINUDA experiment at DAΦNE is bringing a completely new experience and fresh data. If operations prove successful as hoped, there will be the strong case for asking a luminosity up to $10^{34}$cm$^{-2}$sec$^{-1}$ to start looking systematically for neutron rich hypernuclei, and to study hypernuclear transitions with the emission of hard photons. This program is complementary to the JPARC facility one, under construction in JAPAN.

## 3. LOW ENERGY FACTORIES ISSUES

The accelerator group was attended in majority by the DAΦNE community, with some participation from the other lepton factories.

Status reports on existing and in construction factories in the low energy regime were reported: the Beijing τ-charm, the Novosibirsk VEPP-2000, the CESR-c evolution and DAΦNE.

The energy scaling of many parameters of the accelerator was a common argument. Specially all phenomena dominated by synchrotron radiation and impedance, and their impact on the beam-beam interaction, have been discussed.

The round-beam collisions in VEPP-2000 are expected for the end of next year. This collider, the first one built based on the principle of colliding round beams, will show the feasibility of very high beam-beam tune shifts.

The τ-charm factory in Beijing is based on the exploitation of the present BEPC ring, which is used also for synchrotron radiation experiments, plus the construction of a second ring. Double symmetric rings, with one Interaction Region and multibunch flat beams is the design basis for the factory. It is already in the construction stage and the first beam is foreseen in 2007, when CESR-c will be shutdown. CESR-c, which is running now at lower energy, has installed wigglers to increase radiation damping and a new Interaction Region.

DAΦNE is now running for the third experiment, FINUDA; at the workshop the modifications done to the ring in order to install the experiment, plus the shimming of the wiggler poles in order to optimize the sextupolar components and the dynamic aperture were presented. The original physics program of DAΦNE should be completed in 2-3 years and it was discussed which future is expected for the collider. Two possibilities are envisaged: transform the collider in a light–quark factory doubling its energy or upgrade it to a super-regime factory with luminosity of at least $10^{34}$cm$^{-2}$sec$^{-1}$. This second possibility, much more appealing for its challenge and interest was the main argument of the accelerator working group.

## 4. DAFNE2

One of the possible upgrades of DAΦNE is the increase in energy by at least a factor 2. The energy increase needs changes of some of the present systems: essentially the dipoles, the Interaction Region and the injection. All the other systems, RF, feedback, vacuum, quadrupoles, sextupoles, correction coils, diagnostics, are already dimensioned for the higher energy. The requested luminosity is similar to the one already achieved in DAΦNE at the lowest energy, $10^{32}$ cm$^{-2}$sec$^{-1}$. The project for this upgrade is named DAFNE2, where the 'F' stays for Frascati and the '2' for the c.m. energy in GeV.

The dipole preliminary designs, fitting the present vacuum chambers, are based on the use of two materials, steel and permendur. This last one, having a higher saturation field, will be used on the pole in order to increase the magnetic field on the beam axis, up to the necessary 2.2 T.

The luminosity requirements can be achieved with a total current of 0.5 A in 30 bunches. The corresponding beam-beam tune shifts are $\xi_x / \xi_y = 0.014 / 0.024$, below the limit already achieved in DAΦNE.

The IR design, based on the same principles of the present DAΦNE ones, could use superconducting quadrupoles very similar to those presented by the CESR group.

Two possibilities are being considered for the injection: a Linac upgrade to 1 GeV for on-energy injection, or the collider ramping, the first one optimizes the average luminosity, while the second one optimizes the costs.

## 5. DAΦNE-II

The DAΦNE group has presented some different options for the very high luminosity regime at the Φ-energy.

One is based on the exploitation of collisions of higher energy rings with a very large crossing angle, with the advantage of simplifying all the single ring challenges for low-energy/high-currents regimes. The design of a specific detector for the boosted production of collisions presents some disadvantages with respect to the non-boosted case, and since the luminosity reachable with this scheme is also affected by the geometrical reduction due to the crossing angle, the idea has not been further investigated.

The second option, on which a preliminary conceptual design has been presented named DAΦNE-II, is based on the strong RF focusing principle, very high radiation emission and negative momentum compaction ring configuration.

The strong RF focusing is a modulation of the bunch length along the ring, obtained by a very large longitudinal phase advance, corresponding to synchrotron tune near the half integer. The minimum of the bunch length occurs at the IP, while the maximum occurs at the RF cavity position. Such a scheme needs a large absolute value of the momentum compaction and a high RF voltage, and the ring acts as a magnetic compressor. The high radiation emission is obtained with a lattice based on cells with positive and negative dipoles, in which the damping time is of the order of few msec, about a factor 5 less than the present DAΦNE one.

The dispersion self solution oscillates around zero and in each dipole has the sign opposite to the bending angle and has a high value, so that its contribution to the momentum compaction is always negative and high, almost an order of magnitude with respect to the present DAΦNE one. The advantages of the negative momentum compaction regime, like shorter bunches, and more regular bunch shape are therefore included in the design.

An RF system at 500 MHz, with voltage of the order of 10 MV, for a 100m long ring, and momentum compaction near -0.2, will give bunch lengths at the IP near 2 mm.

The Interaction Region fitting the existing KLOE detector is based on low-beta quadrupoles very close to the Interaction Point, to focus the vertical beta to few millimeters.

Considerations on beam lifetime, background, and dynamic aperture are in the preliminary stage. A design for the normal conducting dipoles at 1.8T was presented.

A preliminary configuration of the feedback system in the regime of very high synchrotron frequency, based on GPboards, was also presented by the PEP-II participants.

Figures 1 and 2 show the betatron functions and the dispersion function respectively along the ring.

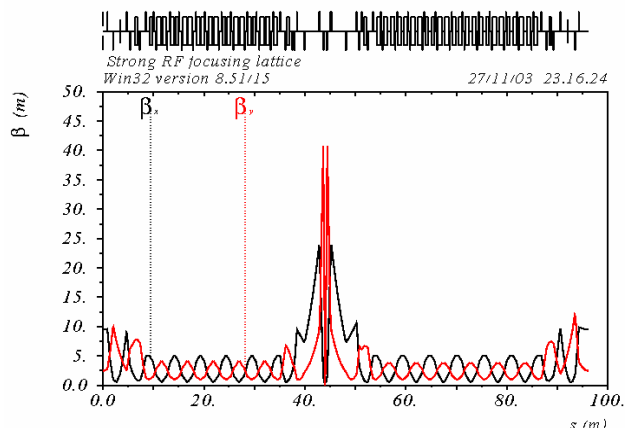

Figure 1 - DAΦNE-II betatron functions of the whole ring.

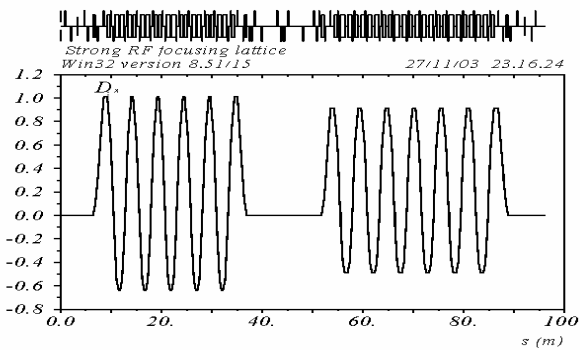

Figure 2 - DAΦNE-II dispersion function along the ring

**WEP02**

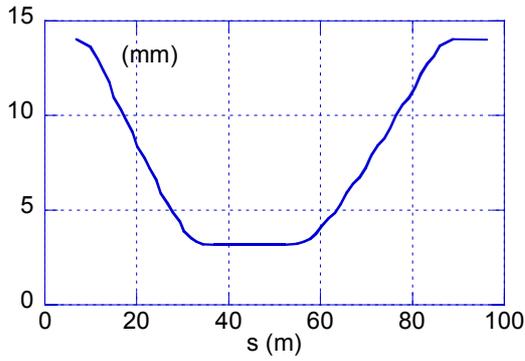

Figure 3 - DAΦNE-II Bunch length along the ring

Table I - DAΦNE-II main parameters

| Parameters | **DAΦNE - II** |
|---|---|
| $E$ (GeV) | .51 |
| $C$ (m) | 96.1 |
| $L$ ($10^{32}$ cm$^{-2}$s$^{-1}$) | 100 |
| # IPs | 1 |
| $\beta^*$ (m) (h / v) | 0.5 / 0.003 |
| $\varepsilon$ (μ rad) (h / v) | 0.3 / 0.003 |
| $\theta$ (mrad) | ± 30 |
| $\sigma_z$ (cm) | 0.3-1.4 |
| $\sigma_E/E$ | 1.3 $10^{-3}$ |
| $N_b$ ($10^{10}$) | 5 |
| $\xi$ (h / v) | 0.06 / 0.05 |
| N bunches | 150 |
| I (A) | 3.4 |
| $f_{RF}$ (MHz) | 499 |
| V (MV) | 9 |
| $\alpha_c$ | -0.2 |
| # dipoles | 44 |
| B dipoles (T) | 1.8 |
| # quadrupoles | 66 |
| $k_1L$ max (m$^{-1}$) | 1 |
| # sextupoles | 22 |
| $k_2L$ max (m$^{-2}$) | 10 |

The bunch length modulation is shown in figure 3, for the set of the collider parameters listed in Table I. These are one of the possible combinations which on paper corresponds to values of luminosity of the order of $10^{34}$ cm$^{-2}$ sec$^{-1}$.

The collider layout fits the existing DAΦNE hall (see fig.4) and could utilize all existing infrastructures and sub-systems. The collider will have only one Interaction Region, while the opposite section will be used for the injection and the RF system. The elements introducing impedance should be placed in the sections where the bunch length is longer, so that perturbation to the bunch distribution is minimized.

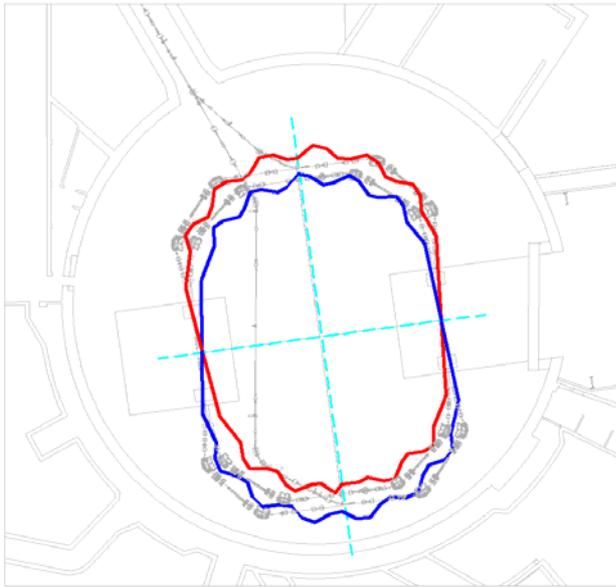

Figure 4: Layout of DAΦNE-II rings inside the present DAΦNE hall

Since the foreseen lifetime at the very high luminosity regime is few minutes, an upgrade of the injection system allowing continuous injection has been defined, in order to increase the average luminosity,. It is based on the doubling of the transfer lines between the damping ring and the collider, while presently the only existing line is shared by both beams, which increases the necessary time for injection.

### 6. CONCLUSIONS

Physics measurement in the 1-2 GeV range will be competitive in next decade, provided that luminosity increases by at least one order of magnitude with respect to what has been obtained up to now.

The overview of e+e- colliders in the low energy range stressed that the evolution of already assessed principles is the next future of many of the present factories (BEPCII, CESRc, DAFNE). There is expectation for the first round colliding beams at VEPP2000.

New ideas for increasing the luminosity are being already tested, i.e. negative momentum compaction. Others are being investigated, and one of the more interesting is the strong RF focusing, which leads to very challenging collider designs.

### References

[1]   http://www.lnf.infn.it/conference/d2